\documentclass[preprint,5p,number,sort&compress]{elsarticle}

\usepackage[utf8]{inputenc}

\usepackage{siunitx}
\DeclareSIUnit\bar{bar}
\usepackage{amsmath}

\begin{document}

\title{Ablation Holes in Tape Targets Induced by Ultra-Intense Laser Pulses}

\author[1]{Michael Ehret\corref{cor1}}
\ead{mehret@clpu.es}
\author[1]{Jon Imanol Api\~{n}aniz}
\author[1]{Jose Luis Henares}
\author[1]{Roberto Lera}
\author[1]{\\Diego de Luis}
\author[1]{José Antonio Pérez-Hernández}
\author[1]{Luca Volpe}
\author[1]{Giancarlo Gatti}

\cortext[cor1]{Corresponding author}

\affiliation[1]{organization={Centro de Laseres Pulsados},
addressline={C/ Adaja 8},
postcode={37185},
city={Villamayor},
country={Spain}}

\begin{abstract}
We present a theoretical modelling able to predict the dimensions of \si{\milli\metre}-sized through-holes produced in interactions of the \SI{1}{\peta\watt} high-power \SI{30}{\femto\second} Ti:Sa laser VEGA-3 with tape targets. We find that sizes of through-holes can be calculated by assuming that the full heat transferred from the laser-heated electron population to the target by electron-electron collisions drives the evaporation of target material. We demonstrate the good reproduction of experimental results.
\end{abstract}



\maketitle

\section{Introduction}

Advances in laser technology \cite{MAIMAN1966,DiDomenico1966,Maine1988,Aoyama2003} make accessible the relativistic interaction regime at PW-class Ti:Sa laser systems opening up to laser driven sources of charged particle beams \cite{TajimaMalka2020} and X-rays \cite{Chakera2008}. Bright ion beams can be generated from solid-density targets by well known mechanisms in the vast field of laser-plasma acceleration \cite{Borghesi2019} such as Target Normal Sheath Acceleration (TNSA) and Radiation Pressure Acceleration. Many applications in science and technology can benefit from laser-driven ion beams such as hadron-therapy \cite{Dosanjh2007,Bulanov2014}, isotope production \cite{Nemoto2001}, positron emission tomography \cite{Santala2001}, ion beam microscopy \cite{Merrill2009}, particle induced X-ray emission \cite{Mirani2021} as well as inertial confinement fusion \cite{Roth2001}.

A major requirement for several of these applications is a high repetition rate (HRR) delivery \SI{>1}{\hertz} of the laser-driven source. Difficulty arises as well studied solid density targets are destroyed by the interaction with ultra-intense high-power laser pulses. The frequent replacement of the target is a hot topic in HRR targetry, e.g. by generation of cryogenic ribbons \cite{Garcia2014}, use of liquid jets \cite{Morrison2018} or by unwinding of a thin tape under the interaction position \cite{Nayuki2003,Noaman2017}.

This study regards the characterization of laser-induced ablation of material from solid tape targets -- for cases in which the extent is so severe that \si{\milli\metre}-sized through-holes are drilled into the target. So far and to our best knowledge, dimensions of tape targets are deduced in heuristic manner by trial and error. We present this modelling of the creation of ablation holes in large solid targets to enable a-priori studies on tape systems.

\section{Material and methods}

Experiments for this work are conducted in the VEGA-3 laser facility at the Centro de Laseres Pulsados (CLPU). Here the high power laser pulse is focused via a $f/10$ off-axis parabola to spots of $d_\mathrm{L}~=~\SI{12}{\micro\metre}$ full-width at half-maximum (FWHM). The pulse is amplified to an energy $E_\mathrm{L}$ up to \SI{30}{\joule}, and compressed to a duration $\tau_\mathrm{L}$ down to \SI{28}{\femto\second}. Only laser pulse energy and duration are varied for this study. After compression, the short laser pulse is transported in high vacuum of \SI{1e-6}{\milli\bar} and the laser focal spot is aligned onto a target for the generation of secondary sources. Routinely reached intensities $I_\mathrm{L} = E_\mathrm{L} / (0.25 \cdot \pi d_\mathrm{L}^2)$ range from \SIrange{1e18}{2e21}{\watt\per\square\centi\metre}. The energy on target is extrapolated from throughput calibrations recorded at low-energy, the focal spot at high energy is estimated to be the same as for low-energy measurements. 

This study uses a tape target, which is a \SI{9}{\micro\metre} thick solid density Al stripe of \SI{10}{\milli\metre} width. The tape normal is vertically inclined by \SI{12.5}{\degree} with respect to the laser axis. A slider frame with \SI{3}{\milli\metre} aperture ensures the tape to stay in the laser-focus plane.

In order to understand the laser-driven heating of the target, numerical simulations are performed for the evolution of the hot electron distribution generated by laser. We follow a detailed model of target charging in short laser pulse interactions \cite{Poye2018} that intends to predict the expected discharge current due to laser-heated relativistic electrons. The numerical code ChoCoLaT2 \cite{Poye2015c} simulates the distribution of electrons generated by laser heating on a thin disk target. The successive electron escape is mitigated by the target potential, based on the driver laser parameters and the interaction geometry. The model takes into account the collisional cooling of electrons within cold solid density targets. The energy and time depending hot electron distribution function $f(E,t)$ describes electrons inside the target and evolves according to
\begin{align}
\partial_t f (E,t) &= \frac{h_\mathrm{L}(E) \Theta(\tau_\mathrm{L} - t)}{\tau_\mathrm{L}}  -  \frac{f (E,t)}{\tau_\mathrm{ee}(E)} - g(E,t) \label{eq:distributionevolution}\\
h_\mathrm{L}(E) &\overset{!}{=} \frac{N_0}{T_0} \exp{\left[ - E/T_0 \right]} \\
N_0 &\overset{!}{=} \int f (E,0) \text{~d}E
\end{align}

\noindent where $h_\mathrm{L}(E)$ is a constant exponential source of hot electrons, $\Theta (t)$ the Heaviside function limiting electron heating to the laser duration, $\tau_\mathrm{ee}(E)$ the energy dependent cooling time and $g(E,t)$ the rate of electron ejection from the target. The initial hot electron temperature $T_0$ depends on laser wavelength and pulse intensity \cite{Fabro1985,Beg1997,Wilks1992}; and $N_0$ is re-normalized to the energy balance $N_0 T_0 = \eta E_\mathrm{L}$ between the total energy of hot electrons in the target and the absorbed laser energy. Simulations require the conversion efficiency $\eta$ of laser energy to energy in the hot electron distribution: for this work a module is added to the code which calculates them automatically according to \cite{Yu1999,Key1998}
\begin{equation}
\eta = 1.2 \times 10^{15} \cdot \left( { I_\mathrm{L} \left[ { \si{\watt\per\square\centi\metre} } \right]  } \right)^{0.75} \quad ,
\end{equation}

\noindent a scaling which is successfully applied to the interpretation of experimental data in the intensity regime from \SIrange{1e18}{1e21}{\watt\per\square\centi\metre} \cite{Schreiber2006,Dover2020}.

The hot electron cooling time depends on target material properties such as mass density $\rho_\mathrm{t}$, mass number $A_\mathrm{t}$, atomic number $Z_\mathrm{t}$, and the hot electron
energy distribution that allows to calculate average speed $\left< v \right>_\mathrm{e}$ and energy $\left< E \right>_\mathrm{e}$. Its meticulous calculation is demonstrated in \cite{Poye2015} with an emphasis on cases relevant for this work. For studies hereinafter, Fe and Ti are added to the library of ChoCoLaT2. 

With further slight modifications to the source code, we extract the branching ratio between energy that leaves the target (transported by escaping electrons) and energy that remains in the target (by collisional damping). A section is added to integrate the the second term of Eq.~(\ref{eq:distributionevolution}) over the full energy distribution. The cumulative sum over all previous evaluation steps $\Delta E_\mathrm{ee} (t)$ is then added to the output generated by the simulation for every recorded time-step.

\section{Theory and calculation}

The energy of the focused laser pulse is partially absorbed by target electrons, which in turn transfer a fraction of their energy to the target bulk material in collisions with cold bulk electrons. If the energy needed to evaporate the volume under the heated area is smaller than the energy transferred, a through-hole can be produced.

The deposited heat is assumed equal to the energy deposited by laser-heated electrons in collisions $\Delta E_\mathrm{ee}$, with efficiency
\begin{equation}
\zeta = \frac{\Delta E_\mathrm{ee}}{\eta E_\mathrm{L}} \quad .
\end{equation}

The \si{\pico\second}-scale duration of the energy transfer from laser to bulk material is much shorter than the timescale of heat diffusion in metals. Therefore we assume the heating process to be instantaneous. Further it is assumed that all the deposited energy is consumed for the evaporation of a cylindrical volume of target material which includes heating, melting, and further heating processes. For a cylindrical volume with radius $R_\mathrm{t}$ and height $ h_\mathrm{t}$, the energy balance equation reads
\begin{multline}\label{eq:model}
\zeta \eta E_\mathrm{L} = \left( h_\mathrm{t} \cdot \pi R_\mathrm{t}^2 \cdot \rho_\mathrm{t} \right) \cdot \\ \left( \int_{T_0}^{T_\mathrm{f}} C_\mathrm{s} \mathrm{\,d}T + S_\mathrm{t} + \int_{T_\mathrm{f}}^{T_\mathrm{b}} C_\mathrm{s} \mathrm{\,d}T + E_\mathrm{t} \right) \quad ,
\end{multline}

where $ \rho_\mathrm{t}$ is the mass density of the tape material. $C_\mathrm{s}$ denotes the specific heat of the target material, $S_\mathrm{t}$ is the latent heat of fusion at melting temperature $T_\mathrm{f}$ and $E_\mathrm{t}$ is the enthalpy of evaporation at the boiling temperature $T_\mathrm{b}$.

\section{Results}

\begin{figure}[htb]
\centering
\includegraphics[width=\linewidth,trim={0mm 60mm 0mm 60mm},clip]{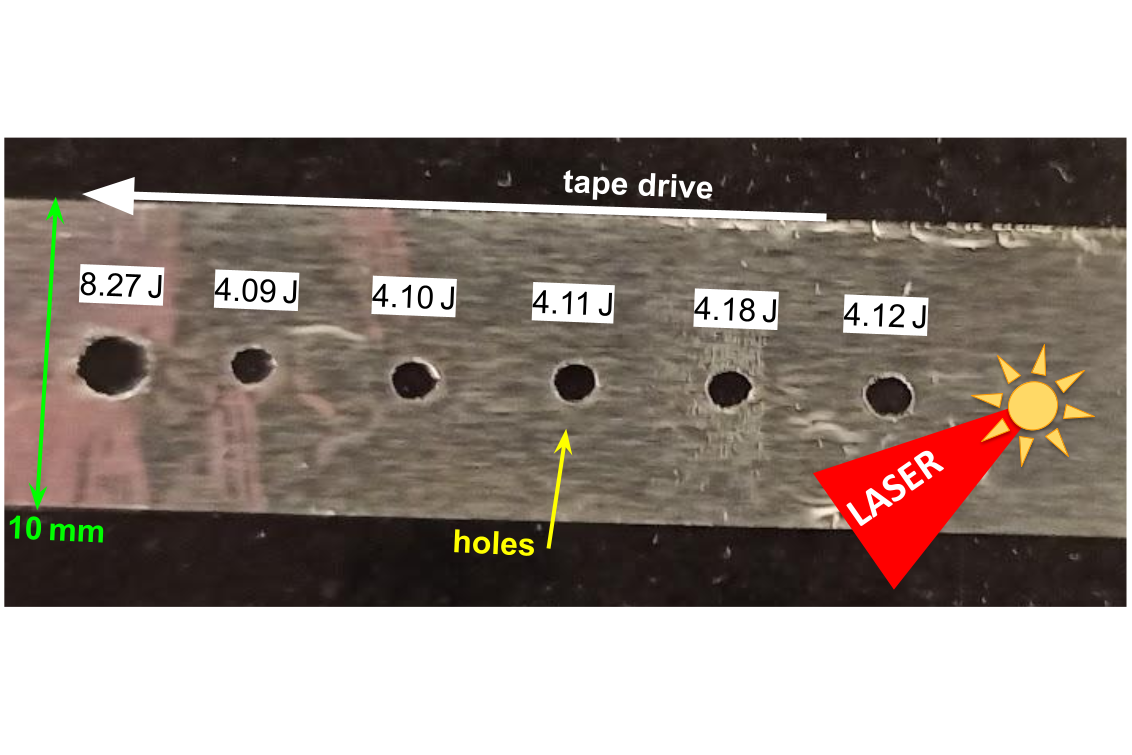}
\caption{Tape target with ablation holes (allow view to black background) and rims of craters (visible through brighter reflectivity of ambient light). Nominal pulse energies range from \SIrange{4}{9}{\joule}, of which \SI{87}{\percent} arrive on-target.}
\label{fig:exresult_2022}
\end{figure}

Tape sections which are shot by laser show clear circular holes opened up by evaporation of target material and rims of ablation craters which can be produced by melting and re-condensation, see fig  \ref{fig:exresult_2022}. The tape drive moves at speeds of the order of \si{\centi\metre\per\second} and does not influence the shapes of holes, there is no systematic ellipticity.

\begin{figure}[htb]
\centering
\includegraphics[width=\linewidth]{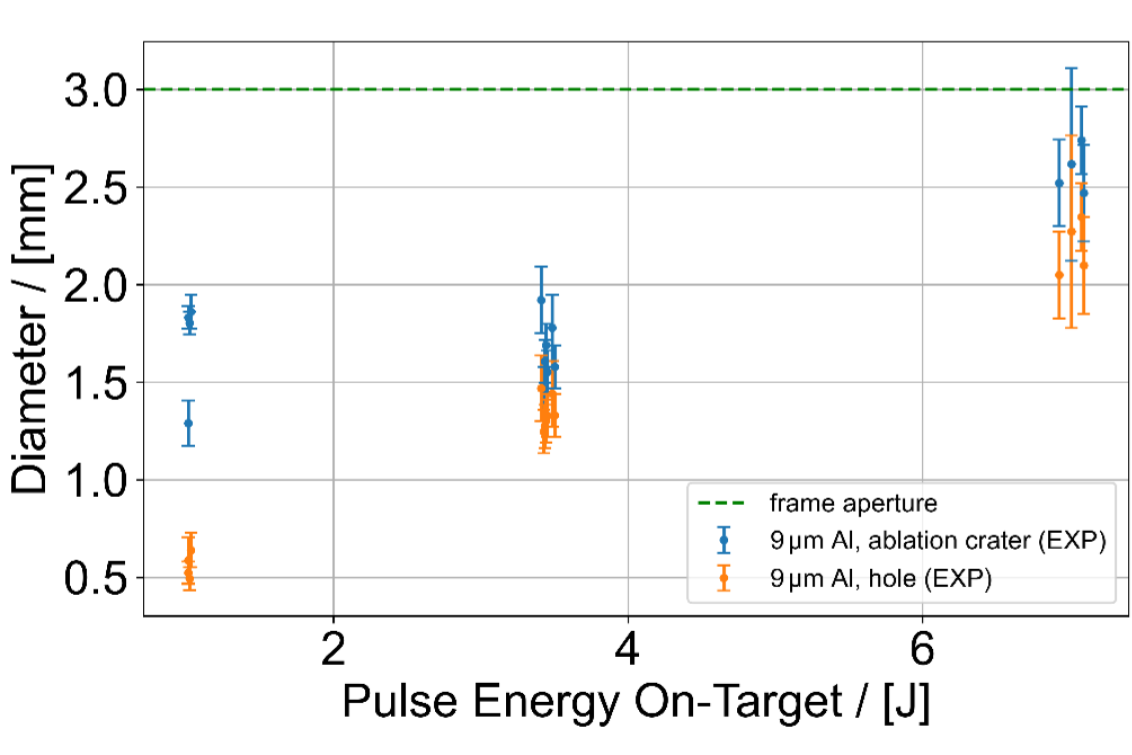}
\caption{The diameter of the evaporation hole under variation of the laser pulse energy at best focus.}
\label{fig:diameterVSpulseenergy_experiment}
\end{figure}

The hole diameters scales with the on-target energy of laser pulses, as shown in Fig.~\ref{fig:diameterVSpulseenergy_experiment}. Compared are shots of different pulse duration but the same focal spot size. With only \si{\micro\metre} thick targets, material from large \si{\milli\metre}-sized areas is evaporated. The variation of results at different pulse duration is below the standard deviation of hole diameters. The later is deduced for each shot from the measurement uncertainty and the difference of the exact hole shapes from perfect circles. Shallow ablation craters are systematically larger than holes, but their sizes approach each other towards higher on-target energy.

\begin{figure}[htb]
\centering
\includegraphics[width=\linewidth]{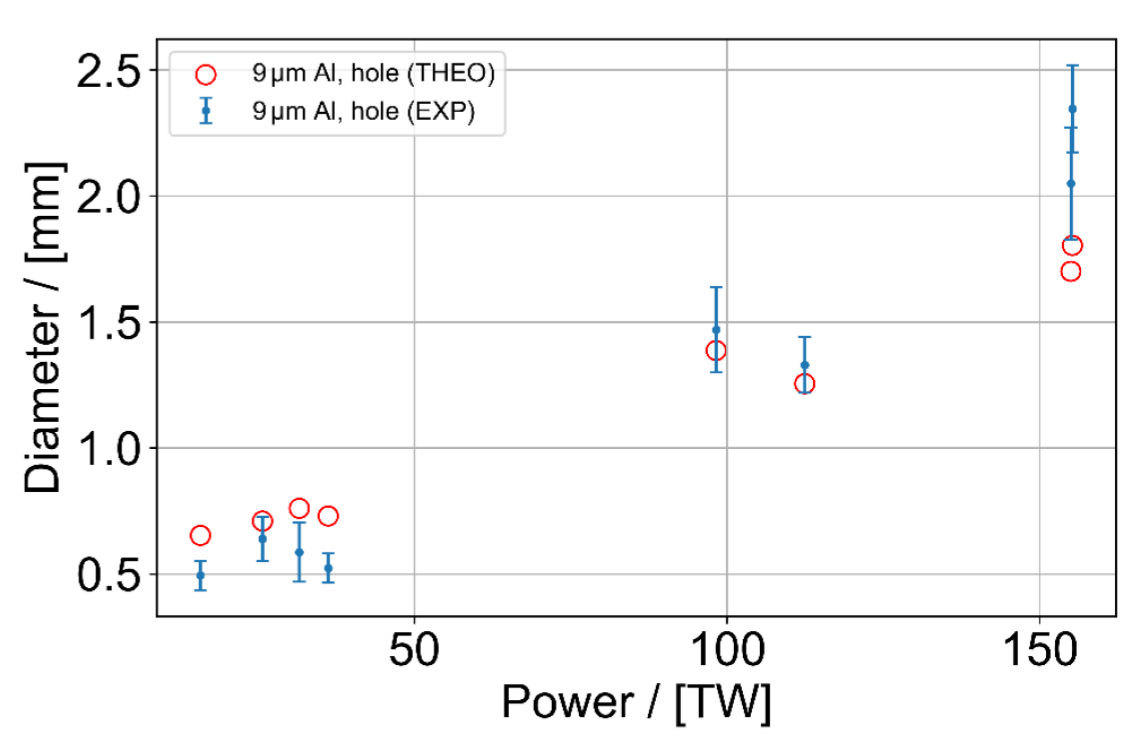}
\caption{The diameter of the evaporation hole under variation of the laser pulse power at best focus, compared are experimental results (EXP) and theoretical predictions (THEO).}
\label{fig:diameterVSpower_modelVSexperiment}
\end{figure}

Our modelling with Eq.~(\ref{eq:model}) allows to calculate the diameters of holes by solving the equation for $R_\mathrm{t}$. As all shots are conducted at best focus, we compare experimental results and theoretical predictions in terms of laser pulse power versus hole diameters in Fig.~\ref{fig:diameterVSpower_modelVSexperiment}. The theoretical prediction is in the margin of uncertainty for \SI{37.5}{\percent} of the data. With shots at a moderate power of \SI{\approx 150}{\tera\watt} each shot evaporates \SI{\approx 0.4}{\milli\gram} of Al. HRR operation at \SI{10}{\hertz} would cause debris of \SI{1}{\kilo\gram} of Al within \SI{72}{\hour}. 

\section{Discussion}

\begin{figure}[htb]
\centering
\includegraphics[width=\linewidth]{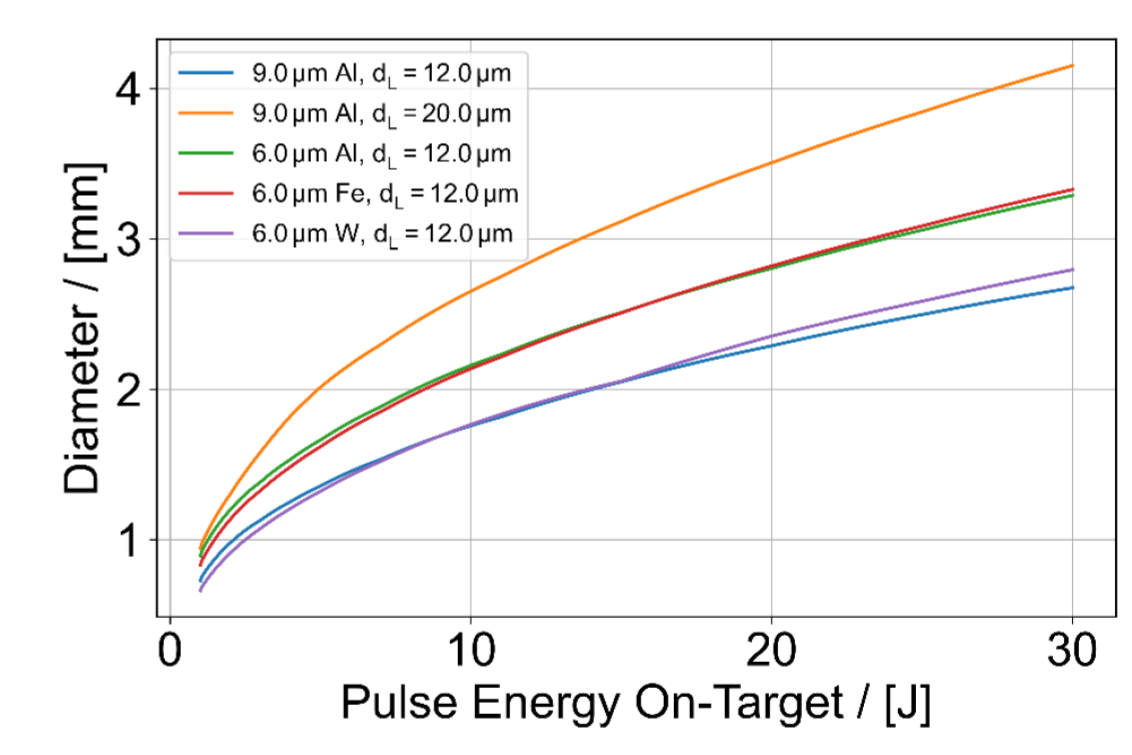}
\caption{The diameter of the evaporation hole under variation of the laser pulse energy at different FWHM focal diameters $d_\mathrm{L}$ with \SI{30}{\femto\second} pulses, and the target characteristics for \SI{10}{\milli\metre} wide tapes of differing thickness.}
\label{fig:2022_res_diameterVSpulseenergy_model}
\end{figure}

Our model allows to deduce theoretical predictions for other laser characteristics, materials and target dimensions. A set of common target materials for laser-accelerators based on TNSA is compared in Fig.~\ref{fig:2022_res_diameterVSpulseenergy_model}. Note that the deposition of energy depends strongly on the laser absorption into hot electrons. The absorption modelling that we apply has a discontinuity at intensities of \SI{3.1e19}{\watt\per\square\centi\metre} \cite{Yu1999,Key1998}, remaining as discontinuity in our model. This is the reason for a small kink in the graphs.

Thinner tapes deliver less areal density to the interaction region and suffer from larger hole diameters. A \SI{6}{\micro\metre} Al tape shows much larger holes compared to a \SI{9}{\micro\metre} Al tape. Despite the larger mass density yielding a higher energy absorption, Fe tapes show smaller hole diameters when compared to Al tapes of similar characteristics. This is due to the larger melting and evaporation temperatures respectively.

Hole diameters can be reduced by choosing a tape material with high melting and evaporation temperature as well with high latent heats, e.g. W. A \SI{6}{\micro\metre} W tape shows similar holes to a \SI{9}{\micro\metre} Al tape. The disadvantage of W with respect to Al is its high mass density, leading to high rates of hot-electron cooling by collisions, thus a lower ejection of relativistic electrons and therefore to a source of lower ion energies at lower number density.

The minimum speed with which the tape should be unwound calculates to $v_\mathrm{tape} = m_0 \cdot 2 R_\mathrm{t}$ where $m_0$ denotes a factor which takes into account the mechanical stability of the punctured tape. It is left to future studies to evaluate this factor in detail for different tape materials. Similar relations will be found for the tape width $w_\mathrm{tape} = m_1 \cdot 2 R_\mathrm{t}$ and the stabilising aperture which guides the tape $a_\mathrm{tape} = m_2 \cdot 2 R_\mathrm{t}$. 

\section{Conclusions}

Our modelling allows to study future tape target materials based on their design parameters without the need for costly experimental campaigns. New materials should have high melting and evaporation temperatures, heat of fusion and evaporation enthalpy respectively. In parallel, a small electron electron collision frequency is beneficial for a small heat supply to the target material. An important parameter for the later is a small mass density of the material.

For stability of the tape, a priori (a) the unwinding speed can now be adjusted in order to be larger than expected hole diameters, (b) the tape width can be chosen much larger than the hole diameter to avoid rupture, and (c) the guiding aperture can be designed slightly larger than holes in order to avoid welding of the tape to the mounting -- thus rupture. Precisely estimated targets make it possible to attach secondary functional structures, e.g. grounding pins. 

Further research effort is required but this model can help to understand the laser induced matter ejection mechanisms in thin foils and high intensities. For the present data sets, the assumption that all the heat stored in hot electron population is spent in evaporation in thermal equilibrium is leading to good agreement with experimental results. Other out-of-equilibrium matter ejection mechanisms like phase explosion \cite{Bulgakova2001} are not taken into account and the agreement is acceptable. New studies with materials of very different thermodynamic parameters (specific heat, latent heat of evaporation, heat conductivity, etc..) will help to further define the model validity.  

\section*{Acknowledgements}

This work would not have been possible without the help of the laser- and the engineering team at CLPU. This work received funding from the European Union’s Horizon 2020 research and innovation program through the European IMPULSE project under grant agreement No 871161 and from LASERLAB-EUROPE V under grant agreement No 871124. It benefited from funding from the Ministerio de Ciencia, Innovación y Universidades in Spain through ICTS Equipment grant No EQC2018-005230-P, further from grant PID2021-125389OA-I00 funded by MCIN / AEI / 10.13039/501100011033 / FEDER, UE and by “ERDF A way of making Europe”, by the “European Union” and in addition from grants of the Junta de Castilla y León with No CLP263P20 and No CLP087U16.

.

\section*{Data statement}

The raw data and numerical methods that support the findings of this study are available from the corresponding author upon reasonable request.


\end{document}